\def\l#1#2{\raisebox{.2ex}{$\displaystyle 
  \mathop{#1}^{{\scriptstyle #2}\rightarrow}$}}
\def\r#1#2{\raisebox{.2ex}{$\displaystyle 
 \mathop{#1}^{\leftarrow {\scriptstyle #2}}$}}

\makeatletter
\def\eqnarray{\stepcounter{equation}\let\@currentlabel=\theequation
\global\@eqnswtrue
\global\@eqcnt\z@\tabskip\@centering\let\\=\@eqncr
$$\halign to \displaywidth\bgroup\@eqnsel\hskip\@centering
  $\displaystyle\tabskip\z@{##}$&\global\@eqcnt\@ne
  \hfil$\displaystyle{\hbox{}##\hbox{}}$\hfil
  &\global\@eqcnt\tw@ $\displaystyle\tabskip\z@
  {##}$\hfil\tabskip\@centering&\llap{##}\tabskip\z@\cr}
\@addtoreset{equation}{section}
  \def\theequation{\thesection.\arabic{equation}}
\makeatother

\documentstyle[12pt]{article}

\begin{document}

\title{Two-dimensional Ultra-Toda integrable mappings and chains
(Abelian case)}

\author{A.N.Leznov \\{\small \it Institute 
for High Energy Physics, 142284 Protvino, Moscow Region, Russia}}

\date{}

\maketitle

\begin{abstract}

The new class of integrable mappings and chains is introduced. Corresponding 
$(1+2)$ integrable systems invariant with respect to such discrete 
transformations are represented in explicit form.
Soliton like solutions of them are represented in terms of
matrix elements of fundamental representations of semisimple
$A_n$ algebras for a given group element.

\end{abstract}

\section{Introduction: the key role of Toda chains in the theory
of integrable systems}

There are many different forms for representation of infinite
Toda chain. The most known and useful are the following ones
\begin{equation}
(\ln v)_{xy}={\r {v}{}\over v}-{v\over \l {v}{}},\quad (\ln\theta)_{xy}=
\r {\theta}{}-2\theta+\l {\theta}{} \label{1}
\end{equation}
where $v\equiv v_s, \r {v}{}\equiv v_{s+1}, \l {v}{}\equiv
v_{s-1}$, $s$ natural number and $x,y$ are independent coordinates of the 
problem. 

Equations (\ref{1}) may be considered as definition of some mapping:
the law by help of which two initial functions $v,\l {v}{}$
($\theta, \l {\theta}{}$) are associated with two final ones $\r
{v}{}, v$ $ (\r {\theta}{},\theta)$.

We restrict ourself by the first system (\ref{1}) and rewrite it
in the following equivalent form
\begin{equation}
\r {u}{}={1\over v},\quad \r {v}{}=v(vu-(\ln v)_{xy}) \label{2}
\end{equation}

The remarkable property of the mapping (substitution) (\ref{2}) consists in
its integrability \cite{1},\cite{2}. This mean that corresponding to 
it symmetry equation (arising as variation derivative of the
substitution by itself)
\begin{equation}
\r {U}{}=-{1\over v^2}V,\quad \r {V}{}=v^2 U+(2uv-(\ln v)_{xy})V-v(
{V\over v})_{xy} \label{3}
\end{equation}
possess the sequence of nontrivial solutions \cite{3}. In (\ref{3}) 
it suggests that independent arguments of functions $U,V$ are $u,v$
(from (\ref{2})) and its derivatives on space coordinates up to
the definite order wile $\r {U}{},\r {V}{}$ are the same functions
arguments of which are shifted by help of (\ref{2}).

Each solution of (\ref{3}) may be connected with the integrable
in Liouville sense (infinite number of conservations "laws" in involution)
system of evolution type equations
\begin{equation}
u_t=U,\quad v_t=V \label{4}
\end{equation}
Moreover the last systems are invariant with respect to substitution 
(\ref{2}).

The most part of integrable systems and equations resolved up to now by
different methods (particular by inverse scattering one) are
directly connected with Toda symmetry (\ref{2}) (or its numerous
auto-Backlund transformations) in described above scheme \cite{4}.

Under appropriate boundary conditions infinite Toda chain is
interrupted and over go into integrable finite dimensional system
different classes solution of which it is possible to represent
in explicit form. The most known ones are Toda chain with fixed ends
$v^{-1}_0=v_N=0$ or periodical Toda $v_0=V_N$. In the first case
it is possible to find general solution \cite{5} (depending on necessary
number of arbitrary functions). In the second case the parametric
soliton-like subclass of solutions \cite{6}. The general
solution in this case may be represented in the form of infinite absolutely
convergent series \cite{7}.

General solution of Toda chain with fixed ends in the case $N=2m$ directly is
connected with m-soliton solution of evolution type equations 
(\ref{4}). Namely $ u_m,v_m$ of Toda chain with fixed
ends is exactly m-soliton solution (under some additional restrictions
on arbitrary functions) of evolution type equations (\ref{4}).

So we see that equations of Toda chain place ambivalent role: it
define the form of integrable systems (as solution of its symmetry equation)
and its interrupting version give possibility to find different classes
solutions of such systems in explicit form.

In papers \cite{1},\cite{2} it was assumed that the theory of integrable 
systems is equivalent to the representation theory of the group
of integrable substitutions with respect to which the Toda chain
system is the simplest partial case.

In the present paper we introduce the new class of integrable
mappings and chains. These chains are different from the usual
Toda one by more number of unknown functions in each point of the
lattice. These mappings are integrable and it is possible to
construct the hierarchy of integrable systems each one of which is
invariant with respect to such discrete substitution. General solution
for the case of fixed ends may be represented in terms of matrix elements
of fundamental representations of semisimple $A_n$ algebras (groups).

\section{Integrable chains connected with the graded algebras}

In the paper \cite{8} (these results are literally repeated in corresponding
chapters of monograph \cite{9}) it was
proposed a general method for construction of exactly integrable
systems connected with arbitrary graded (super) algebras. We
will use it below for the case of $A_n$ semisimple seria.

As usual by $X^{\pm}_{\alpha},h_{\alpha}$ we denote the generators 
of simple roots together with corresponding Cartan elements. The
$\pm s$ graded subspaces consist from generators of $A_n$
algebra which can be constructed from the commutators of $s$ simple
roots. The general equation of \cite{8}
\begin{equation}
[\partial_x-\sum_{s=1}^{m_1} A^{-s},\partial_y-(\rho h)-\sum_{s=1}^{m_2} 
A^{+s}]=0\label{5}
\end{equation}
in the case under consideration may be concretized more detail to
represent integrable chains in more observable form.
Generators of $\pm s$ graded subspaces are the following ones
\begin{equation}
Y^{+s}_{\alpha}=[X^+_{\alpha}...[X^+_{\alpha+s-1},X^+_{\alpha+s}]..],\quad
Y^{-s}_{\alpha}=[[..[X^-_{\alpha+s},X^-_{\alpha+s-1}]....X^-_{\alpha}]
\label{GS}
\end{equation}
with obvious commutation relations as a corollary of their definition:
$$
[Y^{-i}_\alpha,Y^{+j}_\beta]=\delta_{\alpha+i,\beta+j}Y^{-i+j}_\alpha-
\delta_{\alpha,\beta}Y^{-i+j}_{\alpha+j},\quad i\leq j
$$
\begin{equation}
[Y^{+i}_\alpha,Y^{+j}_\beta]=-\delta_{\alpha,\beta+j}Y^{i+j}_\alpha+
\delta_{\alpha+i,\beta}Y^{i+j}_\beta \label{CM}
\end{equation}
$$
[Y^{-i}_\alpha,Y^{+i}_\beta]=\delta_{\alpha,\beta}\sum_{s=0}^{i-1}
h_{\alpha+i}
$$
Further
$$
A^{-s}=\sum_{\alpha}Y^{-s}_{\alpha}f^s_{\alpha},\quad 
A^{+s}=\sum_{\beta}Y^{+s}_{\beta}\bar f^s_{\beta}
$$
The arising as a consequence of (\ref{5}) system of equations takes the form
$$
(f^s_{\alpha})_y+(\sum_{k=0}^{s-1}\bar \rho_{\alpha+k})f^s_{\alpha}-
\sum_{k=1}^{m_1-s}(f^{s+k}_{\alpha}\bar f^k_{\alpha+s}-f^{s+k}_{\alpha-k}
\bar f^k_{\alpha-k})=0
$$
\begin{equation}
\bar \rho_{\alpha}=-\rho_{\alpha+1}+2\rho_{\alpha}-\rho_{\alpha-1},\quad
(\rho_{\alpha})_x+\sum_{s=1}^{min(m_1,m_2)}\sum_{k=0}^{s-1}f^s_{\alpha-k}
\bar f^s_{\alpha-k}=0\label{6}
\end{equation}
$$
(\bar f^s_{\alpha})_x-\sum_{k=1}^{m_2-s}(\bar f^{s+k}_{\alpha} f^k_{\alpha+s}-
\bar f^{s+k}_{\alpha-k}f^k_{\alpha-k})=0
$$
We call (\ref{6}) as UToda$(m_1,m_2)$ chain keeping in mind
that usual two-dimensional Toda system for $A_n$ seria corresponds to the
choice $m_1=m_2=1$. The rezults of \cite{8} give garantie that
system (\ref{6}) is exactly integrable and give prescription how
to integrate it. 

\section{UToda (2,2) system and its general solution}

In this case in each point of the chain there are 5 unknown functions
$\rho_{\alpha},f^1_{\alpha},\bar f^1_{\alpha},f^2_{\alpha},\bar f^2_{\alpha}$.
But due to the gauge invariance not all of them are independent
and after introduction gauge invariant values $q_{\alpha}=f^1_{\alpha}
\bar f^1_{\alpha},p_{\alpha}={f^2_{\alpha}\over f^1_{\alpha}f^1_{\alpha+1}},
\bar p_{\alpha}={\bar f^2_{\alpha}\over \bar f^1_{\alpha}\bar f^1_{\alpha+1}}$
we rewrite (\ref{6}) in the chain form with three independent
functions in each point
$$
(\ln p)_y+(\r {q}{2}\r {p}{}-qp)+(\r {q}{}p-\l {q}{}\l {p}{})=0
$$
\begin{equation}
-(\ln q)_{xy}+(\r {q}{}\bar p-\l {q}{}\l {\bar p}{})_y+(\r {q}{}p-
\l {q}{}\l {p}{})_x+\hat K(q+\l {q}{} q \l {p}{} \l {\bar p}{}+q 
\r {q}{} p \bar p)=0\label{7}
\end{equation}
$$
(\ln\bar p)_x+(\r {q}{2}\r {\bar p}{}-q\bar p)+(\r {q}{}\bar p-
\l {q}{}\l {\bar p}{})=0
$$
where $\hat K \theta=\r {\theta}{}-2\theta+\l {\theta}{}$.

In the case $p=\bar p=0$ we come back to usual Toda chain system
(UToda (1,1)). The case $\bar p=0$ ( or equivalent to it $p=0$)
corresponds to UToda (1,2) chain with equations
$$
-(\ln q)_{xy}+(\r {q}{}p-\l {q}{}\l {p}{})_x+\r {q}{}-2q+\l {q}{} =0
$$
\begin{equation}
{}\label{8}
\end{equation}
$$
(\ln p)_y+(\r {q}{2}\r {p}{}-qp)+(\r {q}{} p-\l {q}{}\l {p}{})=0
$$
The last system is interesting by itself, but always can be
considered as reduction of UToda (2,2) chain under definite
choice of arbitrary functions defined it's solution.

Below we represent the general solution of system (\ref{7}) after
rewriting it in more suitable and observable form.

Let us use the following substitutions
$$
p=\exp (\r {s}{}+s),\quad \bar p=\exp (\r {t}{}+t),\quad \theta=q\exp (s+t)
$$
In new variables system (\ref{7}) takes the form
$$
(\exp-s)_y=\r {\theta \exp-t}{}-\l {\theta \exp-t}{},\quad
(\exp-t)_x=\r {\theta \exp-s}{}-\l {\theta \exp-s}{}
$$
$$
-(\ln \theta)_{xy}+\hat K (\theta \exp -(s+t)+\theta \r {\theta}{}+
\theta \l {\theta}{})=0
$$
or finally after identification $p^{(1}=\exp-s$,$\bar p^{(1}=\exp-t$
$$
(p^{(1})_y=\r {\bar p^{(1} \theta}{}-\l {\bar p^{(1} \theta}{},\quad
(\bar p^{(1})_x=\r {p^{(1} \theta}{}-\l {p^{(1} \theta}{}
$$
\begin{equation}
{}\label{9}
\end{equation}
$$
-(\ln \theta)_{xy}+\hat K (p^{(1} \bar p^{(1} \theta+\theta \r {\theta}{}+
\theta \l {\theta}{})=0
$$
Anyone can agree that the last form is more attractive compare
with (\ref{7}), although both are equivalent to each other (at least in
the case of interrupted chain).

To represent the general solution of (\ref{7}) or (\ref{9}) for
us it will be necessary the knowledge of some facts from
\cite{8}. Here we reproduce them in conspective form. Two
equations of S-matrix type are in foundation of the whole construction
$$
(M_+)_y=M_+L_+\equiv M_-(\sum_1^r Y^{+1}_{\alpha}\bar \phi^1_{\alpha}+\sum_1^
{r-1} Y^{+2}_{\beta}\bar \phi^2_{\beta})
$$
\begin{equation}
{}\label{10}
\end{equation}
$$
(M_-)_x=M_-L_-\equiv M_-(\sum_1^r X^-_{\alpha} \phi^1_{\alpha}+\sum_1^{r-1}
Y^{-2}_{\beta} \phi^2_{\beta})
$$
where $r$ is the rank of semisimple algebra, $Y^{\pm 2}$ are
defined by (\ref{6}).

The "Lagrangian" functions $L^{\pm}$ of equations (\ref{10}) are
correspondingly upper and lower triangular matrixes and by these
reasons solutions of (\ref{10}) may be represented in quadratures.

The solution of the problem may be expressed via matrix elements
of the following $A_n$ group element
\begin{equation}
K=\exp (h\Phi) M^{-1}_-M_+\exp-(h\bar \Phi)\equiv m^{-1}_-m_+ \label{11}
\end{equation}
As it follows from its definition groups elements $m_{\pm}$
satisfy the equations:
$$
(m_+)_y=m_+(-(h\bar \Phi)_y+\sum Y^{+1}_j(\bar \nu_{j+1}-\bar \nu_{j-1})+\sum 
Y^{+2}_j\equiv
$$
$$
m_+(\exp (h\bar \Phi) L_+\exp- (h\bar \Phi)-(h\bar \Phi)_y)
$$
\begin{equation}
{}\label{AD}
\end{equation}
$$
(m_-)_x=m_-((h\Phi)_x+\sum Y^{-1}_j(\nu_{j+1}-\nu_{j-1})+\sum  
Y^{-2}_j\equiv 
$$
$$
m_-(\exp (h\Phi) L_-\exp- (h\Phi)-(h\Phi)_x)
$$
The last equalities determine all introduced above values and relations between
them.

By $\parallel i\rangle, (\langle i\parallel)$ we will denote
the minimal vector of i-th fundamental representation of $A_n$ algebra
with the properties 
$$
X^-_{\alpha} \parallel i\rangle=0,\quad h_s \parallel i\rangle=-\delta_{s,i},
\quad \langle i\parallel X^+_{\alpha}=0,\quad \langle i\parallel h_s=
-\delta{s,i}
$$
\begin{equation}
{}\label{HV}
\end{equation}
$$
X^+_{\alpha} \parallel i\rangle=\delta_{\alpha,i}X^+_i \parallel i\rangle,
\quad \langle i\parallel X^-_{\alpha}=\delta_{\alpha,i}\langle i\parallel X^-_i
$$
The following abbreviations will be used throughout the whole paper
$$
[i]=\langle i\parallel K \parallel i\rangle,\quad \theta_i={[i+1][i-1]\over
[i]^2}
$$
$$
\alpha_{ij..l}\equiv{\langle i\parallel X^-_iX^-_j...X^-_l K \parallel i
\rangle\over \langle i\parallel K\parallel i\rangle},\quad
\bar \alpha_{ij..l}\equiv{\langle i\parallel KX^+_l....X^+_jX^+_i \parallel i
\rangle\over \langle i\parallel K\parallel i\rangle}
$$
In these notations the general solution of the system (\ref{9})
may be represented in the form
$$
\bar p^{(1}_i=(\bar \nu _{i+1}-\bar \alpha_{i+1}-\bar \nu_{i-1}+\bar 
\alpha_{i-1}\equiv(\theta_i)^{-1} (\alpha_i)_y
$$
\begin{equation}
\theta_i={[i+1][i-1]\over [i]^2} \label{12}
\end{equation}
$$
\bar p^{(1}_i=(\nu _{i+1}-\alpha_{i+1}-\nu_{i-1}+ \alpha_{i-1}\equiv(\theta_i)
^{-1} (\bar \alpha_i)_x
$$
In"old" variables solution of (\ref{7}) may be expressed via (\ref{12}) as  
$$
q_i=p^{(1}_i\bar p^{(1}_i\theta_i,\quad \bar p_i={1\over \bar p^{(1}_i
\bar p^{(1}_{i+1}},\quad 
p_i={1\over p^{(1}_i p^{(1}_{i+1}}
$$ 
For checking of the validity of represented above solution only one
relation between matrix elements of different fundamental
representations is necessary. Namely
\begin{equation}
Det_2 \pmatrix{ 
\langle i\parallel K \parallel i\rangle & \langle i\parallel X^-_iK \parallel 
i\rangle \cr \langle i\parallel KX^+_i \parallel i\rangle & \langle i\parallel 
X^-_iKX^+_i \parallel i\rangle \cr}=[i+1][i-1] \label{Y}
\end{equation}
In fact (\ref{Y}) is nevertheless then famous Yakoby equality connecting
the determinants of $i,i+1,i-1$ orders rewritten in the more economical form
\cite{5}.

The proving of this relation reader can find in \cite{10} (see
also appendix). All other necessary relations are direct corollary of the 
last one.

By help of these relations it is not difficult to prove that
$$
(\ln [ i ])_{xy}=\theta_i\theta_{i+1}+\theta_i\theta_{i-1}+\bar p^{(1}_i
p^{(1}_i\theta_i
$$
and other equalities of the same kind ( partially containing in
equations of equivalence (\ref{12})). The details of
corresponding calculations reader can find in section 5.

Solution of UToda(1,2) chain contains among constructed
above. By the same kind of transformation as over go from (\ref{7})
to (\ref{9}) we obtain instead of (\ref{8})
\begin{equation}
\phi_y=\r {\theta}{}-\l {\theta}{},\quad (\ln \theta)_{xy}=\r {\phi
\theta}{}-2\phi\theta+\l {\phi\theta}{}\label{13}
\end{equation}
For general solution of the last chain we have: $\phi$ coincides with 
$p^{(1}_i$ of (\ref{12}) and in the expression for $\theta$ 
(\ref{12}) it is necessary a little modification:
$$
\theta_i=\nu_i\bar \phi^1_i {\langle i-1\parallel K \parallel i-1\rangle
\langle i+1\parallel K \parallel i+1\rangle\over (\langle i\parallel K
\parallel i\rangle)^2}
$$
Of course in equation determining $M_{\pm}$ (\ref{10}) it is
necessary put $\bar \phi^2_i=0$.

From explicit form of solution (\ref{12}) we see that it defines
by only one group element $K$ and so it is possible to
wait that all problems connected with UToda chains systems may
be resolved on the level of its properties.

\section{Parameters of evolution - Hamiltonians flows}

In this section we introduce the parameters of evolution and
represent the way of construction the systems of equations invariant
with respect to UToda substitutions. We begin the discussion
from the simplest case of usual Toda chain for which solution of
the problem is known \cite{3}.

\subsection{Two-dimensional Davey-Stewartson hierarchy}

In this case the group element $m_+$ is defined by equation
\begin{equation}
m'_+\equiv(m_+)_y=m_+(-(h\bar \Phi)'+\sum Y^{+1}_j \label{14}
\end{equation}
and depend on the set of arbitrary functions $\bar \Phi_i$. Let
try to find the last functions in such a way that equation
\begin{equation}
\dot m_+\equiv (m_+)_{t_2}=m_+(-\dot {(h\bar \Phi)}+\sum
Y^{+1}_j\mu_j-\sum Y^{+2}_j) \label{15}
\end{equation} 
would be selfconsistent with (\ref{14}). Maurer-Cartan identity
after its trivial resolution takes the following form
\begin{equation}
\mu_i=-\bar \Phi_{i+1}+\bar \Phi_{i-1},\quad \mu'_i+\dot {(k\bar
\Phi)_i}-\mu_i(k\bar \Phi)'=0\label{16}
\end{equation}
where as usually $(kf)_i=-f_{i+1}+2f_i-f_{i-1}$.
Finally (\ref{16}) is equivalent to
\begin{equation}
\dot {\bar \Phi_i}-\dot {\bar \Phi_{i-1}}+\bar \Phi''_i+ \bar \Phi''_{i-1}+
(\bar \Phi'_i-\bar \Phi'_{i-1})^2=0\label{17}
\end{equation}
( in all cases from the condition $b_i=b_{i+1}$ we have done
conclusion $b_i=0$).

Solution of the chain system (\ref{17}) it is possible to express via the $N$
($N$-is the number of the points of interrupted chain) linear independent
solutions of the single one-dimensional Schrodinger equation 
$$
\dot \Psi=\Psi''+V(t_2,y)\Psi
$$
where $V$ is arbitrary function of space $y$ and time $t_2$ coordinates 
\cite{3}.

The chain (ref{17}) induced the Davey-Stewartson system \cite{11}. 
To explain this fact let us consider matrix element $[ i ]$ and calculate its 
derivatives with respect to arguments $y,t_2$ ( we use notations
introduced in (\ref{HV}) and below). As a consequence of
(\ref{14}),(\ref{15}) and (\ref{17}) we have
$$
\dot {\ln [i ]}=\dot {\bar \Phi_i}+
\mu_i \bar \alpha_i-\bar \alpha_{i,i+1}+\bar \alpha_{i,i-1}
$$
\begin{equation}
(\ln [ i ])'=(\bar \Phi_i)'+\bar \alpha_i
\label{18}
\end{equation}
$$
{[ i ]''\over [ i ]}=(\bar \Phi_i)''+(\bar \Phi'_i)^2+(\bar \Phi_{i+1}+
\bar \Phi_{i-1})'\bar \alpha_i+\bar \alpha_{i,i+1}+\bar \alpha_{i,i-1}
$$

Excluding $\dot {\bar \Phi_i},\bar \Phi'_i,\bar \Phi''_i$ from (\ref{17})
by help of (\ref{18}) we come to the key equality
\begin{equation}
\dot {\ln{[ i ]\over [ i-1 ]}}+(\ln ([ i ][ i-1 ])''+((\ln {[ i ]\over 
[ i-1 ]}')^2=2(\bar \alpha_{i-1,i}+\bar \alpha_{i,i-1}-\bar \alpha_{i-1}
\bar \alpha_i) \label{19}
\end{equation}
remarkable by the fact that its right-hand side is identically
equal to zero due to recurrent relations between the matrix
elements of different representations of $A_n$ groups (see appendix).

Introducing the functions $v={[ i ]\over [ i-1 ]},u={[ i-2
]\over [ i-1 ]}$, bearing in mind the main equation of Toda
chain by itself
$$
(\ln [ i ])_{xy}={[ i+1 ][ i-1 ]\over
[ i ]^2}
$$
and the fact that equality (\ref{19}) is correct for arbitrary
$i$, we conclude that functions $u,v$ satisfy the following
system of equations
\begin{equation}
-\dot u+u_{yy}+2u\int dx(uv)_y=0\quad \dot v+v_{yy}+2v\int dx(uv)_y=0
\label{20}
\end{equation}
This is exactly Davey-Stewartson system \cite{11}. In one-dimensional
limit - usual nonlinear Schrodinger equation.

In general case equation (\ref{15}) defined algebra valued function
$m_+^{-1}\dot m_+$ it is necessary to change on the condition that this 
function is decomposed on generators of algebra the graded index of which is 
less then some given natural number say $r$. In this case we will obtain system
of equation which determine dependence $\bar \Phi_i$ on 
parameter $\bar t_r$ and obtain the corresponding system of equations of
two-dimensional D-S hierarchy. By different method this problem
in explicit form was solved in \cite{3}, \cite{10}.

The construction described above in one-dimensional case
equivalent to multi-time formalism and corresponding technique
of Hamiltonian flows \cite{4}.

Of course all done above it is possible to repeat with respect space
coordinate $x$ in the pair with group element $m_-$.

As a result we will obtain the sequence of right $\bar t_s$ and
left $t_l$ evolution parameters, corresponding system of
equation invariant with respect to Toda discrete substitution and
it's particular explicit multi-soliton type solutions.

\subsection{PToda(2.2) case}

Now we want to apply the technique of the last subsection for construction
of unknown up to now example of integrable system invariant with respect
UToda(2,2) substitution (\ref{9}). We omitted as a rule the
calculations by themselves, representing only the finally results. 
All necessary formulae for it's independent verification reader
find in section 5 and Appendix.
These calculation are not difficult but very long in consequent rewriting 
(may be because of the bad notations or very straightforward attempts to 
realize them by known for us methods).

In this case element $m_+$ satisfy the equation (see section 3):
$$
(m_+)'=m_+(-(h\bar \Phi)'+\sum Y^{+1}_j\bar \phi_j+\sum Y^{+2}_j\equiv
$$
$$
m_+(\exp (h\bar \Phi) L_+\exp- (h\bar \Phi)-(h\bar \Phi)')\quad
f'\equiv f_y,\quad\bar \phi_j=(\bar \nu_{j+1}-\bar \nu_{j-1})
$$

Corresponding operator of $t_2$ differentiation has the form
\begin{equation}
\dot m_+=m_+((-\dot {(h\bar \Phi)}+\sum Y^{+1}_j\mu^{(1}_j+
\sum Y^{+2}_j\mu^{(2}_j+\sum Y^{+3}_j\mu^{(3}_j-\sum Y^{+4}_j) \label{21}
\end{equation} 
Condition of selfconsistency (Maurer-Cartan identity) gives possibility
to express all functions $\mu^{(s}_i$ from (\ref{21}) in terms of
$\bar \Phi_i, \bar \nu_i$ and find the system of equations which
the last functions as functions of $y,t_2$ arguments satisfy.

By help of commutation relations (\ref{CM}) all calculations are
straightforward. Below reader can find result of them:
$$
\mu^{(3}_i=\bar \phi_{i+2}+\bar \phi_i=\bar \nu_{i+3}-\bar \nu_{i-1}\quad
\mu^{(2}_i=\bar \Phi'_{i+1}-\bar \Phi'_{i+2}-\bar \Phi'_i+\bar \Phi'_{i-1}+
\bar \phi_i\bar \phi_{i+1}
$$
$$
\mu^{(1}_i=-(\bar \nu_{i+1}-\bar \nu_{i-1})(\bar \Phi'_{i+1}-\bar \Phi'_{i-1})-
(\bar \nu'_{i+1}+\bar \nu'_{i-1})
$$
The chain system of equation with respect to unknown functions
$\bar \Phi,\bar \nu$ ( compare with (\ref{17}) in this case
has the form:
$$
\dot {\bar \Phi}_{i+1}-\dot {\bar \Phi}_{i-1}+\bar \Phi''_{i+1}+\bar \Phi''_
{i-1}+(\bar \Phi'_i-\bar \Phi'_{i+1})^2+(\bar \Phi'_i-\bar \Phi'_{i-1})^2=
2\bar \nu'_i(\bar \nu_{i+1}-\bar \nu_{i-1})
$$
$$
\dot {(\bar \nu_{i+1}-\bar \nu_{i-1})}+(\bar \nu_{i+1}+\bar \nu_{i-1})''-
2\bar \nu'_{i+1}(\bar \Phi'_i-\bar \Phi'_{i-1})-2\bar \nu'_{i-1}(\bar \Phi'_i-
\bar \Phi'_{i+1})+
$$
\begin{equation}
{}(\bar \nu_{i+1}-\bar \nu_{i-1}) \times\label{22}
\end{equation}
$$
(\dot {\bar \Phi}_{i+1}-2\dot {\bar \Phi}_i)+
\dot {\bar \Phi}_{i-1}+\bar \Phi''_{i+1}-\bar \Phi''_{i-1}+(\bar \Phi'_{i+1})^2
-(\bar \Phi'_{i-1})^2-2(\bar \Phi'_{i+1}-\bar \Phi'_{i-1})\bar \Phi'_i=0
$$

In what follows in this section we deviate from introduced in
the last section notations and consider UToda$(2,2)$ in variables
$$
p_i=\alpha_i-\nu_i,\quad \bar p_i=\bar \alpha_i-\bar \nu_i
$$
In this variables as it follows from (\ref{12}) UToda$(2,2)$ substitution
takes the form
$$
(p_i)_y=\theta_i(\bar p_{i-1}-\bar p_{i+1}),\quad (\bar p_i)_x=\theta_i
(p_{i-1}-p_{i+1})
$$
\begin{equation}
(\ln \theta)_{xy}=\hat K (p_y \bar p_x \theta+\theta \r {\theta}{}+
\theta \l {\theta}{})\label{23}
\end{equation}
Let us define functions $v_i={[ i+1 ]\over [ i ]},u_i={[ i-1 ]\over [ i ]}$.
Obviously $\theta_i\equiv u_i v_i,u_{i+1}=v^{-1}_i$.

The following equalities are direct corollary of all introduced above
definitions and may be verified directly (all necessary formulae
reader can find in section 5 and in Appendix)
$$
{\dot v_i\over v_i}+{v''_i\over v_i}+2(\bar \alpha_{i,i-1}-\bar
\nu_{i-1} \bar \alpha_i)'-2\bar p'_i\bar p_{i+1}=V_i(y,t_2)
$$
$$
-{\dot u_i\over u_i}+{u''_i\over u_i}-2(\bar \alpha_{i,i+1}-\bar
\nu_{i+1} \bar \alpha_i)'+2\bar p'_i\bar p_{i-1}=U_i(y,t_2)
$$
$$
\dot p_i=-p_i''-2(\ln v_{i-1})'+2\theta_i\bar p'_{i-1}
$$
$$
V_i=\dot {\bar \Phi}_{i+1}-\dot {\bar \Phi}_i+\bar \Phi''_{i+1}-\bar \Phi''_i
+(\bar \Phi'_i-\bar \Phi'_{i+1})^2-2\bar \nu'_i\bar \nu_{i+1}
$$
$$
U_i=-\dot {\bar \Phi}_{i-1}+\dot {\bar \Phi}_i+\bar \Phi''_{i-1}-\bar \Phi''_i
+(\bar \Phi'_i-\bar \Phi'_{i-1})^2+2\bar \nu'_i\bar \nu_{i-1}
$$
To have some closed system it is necessary to exclude from the
last system of equalities the terms containing  functions $\bar \alpha_{i\pm1},
\bar \alpha _{i,i\pm1}$. The following additional equalities 
$$
(\bar \alpha_{i,i\pm1}-\bar \nu_{i\pm1} \bar \alpha_i)_x=\mp\theta_i
\theta_{i\pm1}+\theta_i\bar p_{i\pm1}(\bar p_i)_x
$$
solve this problem. After this keeping in mind equations of substitution
(\ref{23}) it is possible to rewrite the previous system of
equalities in the closed form for 8 "unknown" functions $v_{i+1},v_i,
u_i,u_{i-1},p_i,p_{i+1},\bar p_i,\bar p_{i-1}$
$$
\dot v_{i+1}+v''_{i+1}+2v_i\bar p'_{i+1}p'_{i+1}+2v_{i+1}\bar p_{i+1}p_i -
2v_{i+1}\int dx[-u_iv_{i+1}+u_iv_i\bar p_{i+1}(p_i)_x]'=0
$$
$$
\dot v_i+v''_i-2v_i\bar p_{i+1}\bar p'_i +
2v_i\int dx[v_iu_{i-1}+u_iv_i\bar p_{i-1}(p_i)_x]'=0
$$
$$
-\dot u_i+u''_i+2u_i\bar p_{i-1}\bar p'_i-
2u_i\int dx[-u_iv_{i+1}+u_iv_i\bar p_{i+1}(p_i)_x]'=0
$$
$$
-\dot u_{i-1}+u''_{i-1}+2u_i\bar p'_{i-1}p'_{i-1}-2u_{i-1}\bar p_{i-1}p'_i +
2v_{i+1}\int dx[v_iu_{i-1}+u_iv_i\bar p_{i-1}(p_i)_x]'=0
$$
\begin{equation}
\dot p_i+p''_i-2(\ln u_i)'p'_i-2u_iv_i \bar p'_{i-1}=0 \label{24}
\end{equation}
$$
\dot p_{i+1}+p''_{i+1}+2(\ln v_i)'p'_{i+1}-2{v_{i+1}\over v_i} \bar p'_i=0
$$
$$
\dot {(\theta^{-1}_i (\bar p_i))_x}=p''_{i+1}+p''_{i-1}+2p'_{i+1}(\ln v_i)'+
2p'_{i-1}(\ln u_i)'-2\bar p'_i (\theta_{i+1}+\theta_{i-1})
$$
$$
\dot {(\theta^{-1}_{i-1} (\bar p_{i-1}))_x}=p''_i+p''_{i-2}+2p'_i(\ln v_{i-1})'
+2p'_{i-2}(\ln u_{i-1})'-2\bar p'_{i-1} (\theta_i+\theta_{i-2})
$$

The last two equations are direct consequence of the two
previous ones by help of UToda $(2,2)$ transformation may be represent
in terms of only unknown functions. So this system is closed,
integrable and sequence of its particular solutions is given by formulae
$$
p_i=\alpha_i-\nu_i,\quad \bar p_i=\bar \alpha_i-\bar \nu_i,\quad
v_i={[i+1]\over [i]},\quad u_i={[i+1]\over [i]}
$$
as it was shown above. 

Now we are able to clarify situation with solution of chain system 
(\ref{22}). It is obvious that the system (\ref{24}) possess
particular solution of the form
$$
u_0=u_{-1}=p_0=\bar p_0=p_{-1}=\bar p_{-1}=0
$$
Indeed in the case of final dimensional algebra $A_n$ all matrix
elements above are equal to 0. For remaining unknown functions $v_0,v_1,p_1,
\bar p_1$ as a corollary of (\ref{24})  we obtain the following (unclosed)
system of equations (after trivial regrouping)
$$
\dot v_0+v''_0=V_0v_0,\quad \dot (p_1 v_0)+(p_1 v_0)''=V_0(p_1 v_),\quad
\dot v_1+v''_1-U_0v_1=-2p'_1\bar p'_1 v_0
$$
(the arising of arbitrary functions $(U_,V_0)$ is connected with
ambiguity of $\int dx 0=F(y,t_2)$).

So we see that the functions are $v_0,p_1 v_0)$ are linear
independent solutions of Schrodinger equation with arbitrary
potential function $V_0$. While $v_1$ is the solution
unhomoginios Schrodinger equation (with known search function $-2p'_1\bar p'_1
v_0$ and potential (also arbitrary function) $U_0$.

In terms of different solutions of tis pair of Schrodiger
equations it is possible represent general solution of the chain
(\ref{22}). To this problem we hope to come back in some other place.

\section{UToda($m_1,m_2$) system and its general solution}

In this section we represent some necessary auxiliary relations
of representation theory of semisimple algebras. Construction
the equations of UToda $(m_1,m_2)$ chains together with its
general solution after this takes the form of pure technical manipulations.

Let the  pair of operators $m^{\pm}$ satisfy the equations, generalizing
(\ref{AD})
$$
m^+_y=m^+(-(h\bar \Phi)_y+\sum_{s=1}^q (Y^s\bar \phi^{(s})\equiv m^+L^+
$$
\begin{equation}
(Y^s\bar \phi^{(s}) \equiv \sum_i (Y^s_i\bar \phi^{(s}_i), \quad
(Y^s \phi^{(s})^T \equiv \sum_i (Y^{-s}_i \phi^{(s}_i) \label{24}
\end{equation}
$$
m^-_x=m^-((h \Phi)_x-\sum_{s=1}^q (Y^s\bar \phi^{(s})^T\equiv m^+L^-
$$
where generators $Y^{\pm s}$ together with their commutation relations
are defined by (\ref{6}) and below.

Our nearest goal is to find recurrent relations (or equations) which
some matrix elements of the different fundamental
representations of the single group element
\begin{equation}
k=m^{-1}_-m_+\label{25}
\end{equation}
satisfy. We have  
\begin{equation}
(\ln [ i ])_{xy}=[ i ]^{-2} \pmatrix{ 
\langle i\parallel K \parallel i\rangle & \langle i\parallel L^-K \parallel 
i\rangle \cr
\langle i\parallel KL^+ \parallel i\rangle & \langle i\parallel L^-KL^+ 
\parallel i\rangle \cr}\label{26}
\end{equation}
In connection with (\ref{HV}) under the action on minimal state
vector by the operators of the simple roots only $X^+_i\parallel i\rangle \neq
0,\langle i\parallel X^-_i \neq 0$. By this reason the action of
operators $L^{\pm}$ on the minimal state vector may be
represented in the form
$$
L^+\parallel i\rangle=l^+_i X^+_i \parallel i\rangle,\quad
\langle i\parallel L^-=\langle i\parallel  X^-_i l^-_i
$$
where $l^{\pm}_i$ some operators polynomials in generators of positive
(negative) simple roots. For instance
$$ 
(Y^2\bar \phi^{(2}) \parallel i\rangle=(\phi^{(2}_{i-1} X^+_{i-1}-\phi^{(2}_
i X^+_{i+1})  X^+_i \parallel i\rangle
$$
or in this case $l^+_i=\phi^{(2}_{i-1} X^+_{i-1}-\phi^{(2}_{i+1} X^+_{i+1}$.
Keeping this fact in mind and taking into account (\ref{6}) we can rewrite 
(\ref{26}) in the form
\begin{equation}
(\ln [ i ])_{xy}=[ i ]^{-2} (l^-_i)_l (l^+_i)_r [ i-1 ][ i+1]\label{27}
\end{equation}
where now $(l^-_i)_l, (l^+_i)_r$ are the same polynomials
constructed from the generators of simple roots correspondingly
of the left and right adjoint representations. Formulae below are
illustration of application of the last general equalities (\ref{27})
to concrete cases under the choice $q=3$ in (\ref{24}).
$$
(\alpha_i)_y=\theta_i (\bar \phi^{(1}_i+
\bar \phi^{(2}_{i-1}\bar \alpha_{i-1}-\bar \phi^{(2}_i\bar \alpha_{i+1}+
\bar \alpha_{i+1,i+2}-\bar \alpha_{i+1}\bar \alpha_{i-1}+\bar \alpha_{i-1,i-2})
$$
\begin{equation}
(\alpha_{i,i+1})_y=\alpha_{i+1}(\alpha_i)_y-\theta_i\theta_{i+1}(\bar 
\phi^{(2}_i+\bar \alpha_{i-1}-\bar \alpha_{i+2})\label{28}
\end{equation}
$$
(\ln [ i ])_{xy}=\theta_{i+2}\theta_{i+1}\theta_i+\theta_{i+1}\theta_i\theta_
{i-1}+\theta_{i-2}\theta_{i-1}\theta_i+
$$
$$
\theta_{i+1}\theta_i(\phi^{(2}_i+\alpha_{i-1}-\alpha_{i+2})\bar {(....)}+
\theta_{i-1}\theta_i(\phi^{(2}_{i-1}+\alpha_{i-2}-\alpha_{i+1})\bar {(....)}
$$
$$
\theta_i (\phi^{(1}_i+\phi^{(2}_{i-1}\alpha_{i-1}-\phi^{(2}_i\alpha_{i+1}+
\alpha_{i+1,i+2}-\alpha_{i+1}\alpha_{i-1}+\alpha_{i-1,i-2})\bar {(....)}
$$
where by $\bar {(....)}$ we understand the same values as in the
first multiplicator in which all functions changed on the bar
ones. The same relations as (\ref{28}) obviously takes place
when all functions $\alpha$ are changed on the bar ones
together with $y\to x$ and visa versa.

Now let us introduce the new functions $p^{(1,2}_i,\bar
p^{(1,2}_i$ by the relations
$$
p^{(1}_i=\theta^{_1}_i  (\bar \alpha_i)_y,\quad \bar p^{(1}_i=\theta^{_1}_i  
(\alpha_i)_x
$$
$$
p^{(2}_i=\phi^{(2}_i+\alpha_{i-1}-\alpha_{i+2},\quad
\bar p^{(2}_i=\bar \phi^{(2}_i+\bar \alpha_{i-1}-\bar \alpha_{i+2}
$$
Using once more (\ref{28}) we obtain the chain of equations
which functions $p^{(1,2}_i,\bar p^{(1,2}_i,\theta_i$ satisfy
$$
(p^{(2}_i)_y=\theta_{i-1} \bar p^{(1}_{i-1}-\theta_{i+2} \bar p^{(1}_{i+2},
\quad (\bar p^{(2}_i)_x=\theta_{i-1} p^{(1}_{i-1}-\theta_{i+2} p^{(1}_{i+2}
$$
$$
(p^{(1}_i)_y=\theta_{i-1} \bar p^{(1}_{i-1}p^{(2}_{i-1}-\theta_{i+1} 
\bar p^{(1}_{i+1}p^{(2}_i+\theta_{i-1}\theta_{i-2}\bar  p^{(2}_{i-2}-
\theta_{i+1}\theta_{i+2} \bar p^{(2}_{i+1}
$$
$$
(\bar p^{(1}_i)_x=\theta_{i-1} p^{(1}_{i-1}\bar p^{(2}_{i-1}-\theta_{i+1} 
 p^{(1}_{i+1}\bar p^{(2}_i+\theta_{i-1}\theta_{i-2} p^{(2}_{i-2}-
\theta_{i+1}\theta_{i+2} p^{(2}_{i+1}
$$
$$
(\ln \theta_i)_{xy}=\hat K (\theta_i\theta_{i+1}\theta_{i+2}+\theta_i\theta_
{i-1}\theta_{i-2}+\theta_i\theta_{i+1} p^{(2}_i\bar p^{(2}_i+\theta_{i-1}
\theta_i p^{(2}_{i-1}\bar p^{(2}_{i-1}+\theta_i p^{(1}_i\bar p^{(1}_i
$$
This is exactly UToda $(3,3)$ chain system with known general solution which
determines by the set of arbitrary functions $(\Phi_i,\phi^{(1}_i, p^{(2}_i)$
and $(\bar \Phi_i,\bar \phi^{(1}_i,\bar p^{(2}_i)$ of single
arguments (x,y) correspondingly.

In general case literally repeating calculations of this section
or corresponding places of sections 2-3 we come finally to
expressions for equtions of integrable UToda $(2,2)$ substitution
$$
(p^{(s}_{\alpha})_y=\sum_{k=1}^{m_1-s}(p^{(s+k}_{\alpha-k}
\bar p^{(k}_{\alpha-k}\prod_{i=1}^k \theta_{\alpha-k+i-1}-p^{(s+k}_{\alpha}\bar p^{(k}_
{\alpha+s}\prod_{i=1}^k \theta_{\alpha+s+i-1})=0
$$
\begin{equation}
(\ln \theta_{\alpha})_{xy}=\hat K \sum_{s=1}^{min(m_1,m_2)}\sum_{k=0}^{s-1}
p^{(s}_{\alpha-k}\bar p^{(s}_{\alpha-k}\prod_{i=1}^s \theta_{\alpha-s+i-1}
,\quad p^{(m_1}_{\alpha}=1,\quad \bar p^{(m_2}_{\alpha}=1\label{29}
\end{equation}
$$
(\bar p^{(s}_{\alpha})_y=\sum_{k=1}^{m_2-s}(\bar p^{(s+k}_{\alpha-k}
p^{(k}_{\alpha-k}\prod_{i=1}^k \theta_{\alpha-k+i-1}-\bar p^{(s+k}_{\alpha} p^{(k}_
{\alpha+s}\prod_{i=1}^k \theta_{\alpha+s+i-1})=0
$$

\section{Outlook}

At first we summarize in few words the construction of the
present paper (excluding all details).

In foundation of it are two groups elements $m_{\pm}$ belonging
correspondingly to $\pm$ resolvable subgroups of some semisimple
group. They determines by the pair of $S$-matrix type equations
\begin{equation}
l_+\equiv m_+^{-1}(m_+)_y=\sum_{s=0}^{m_1} A^{+s},\quad
l_-\equiv m_-^{-1}(m_-)_x=\sum_{s=0}^{m_2} A^{-s} \label{01}
\end{equation}
The nature of these equations are pure algebraic - this is
condition that lagrangian operators are decomposed on the
operators of algebra with graded indexes less then $m_1,(m_2)$.

Matrix elements of different fundamental representations of the
single group element $K=m_-^{-1}m_+$ satisfy definite system of
equalities which can be interpret as as equations of exactly
integrable UToda$ (m_1,m_2)$ system.

On this step construction of UToda $(m_1,m_2)$ integrable
mapping (substitution) is closed.

Next step is connected with introduction of evolution times parameters.
Arbitrary up to now functions $\phi^{(s}_i(y),\phi^{(s}_i(x) $
are restricted by the conditions that elements $m_{\pm}$ satisfy
additional system of equations
\begin{equation}
m_+^{-1}(m_+)_{\bar t_{d_1}}=\sum_{s=0}^{d_1} B^{+s},\quad
m_-^{-1}(m_-)_{t_{d_2}}=\sum_{s=0}^{d_2} B^{-s} \label{02}
\end{equation}
Condition of selfconsistency of (\ref{01}) with (\ref{02})
determines explicit dependence of functions $\phi^{(s}_i\equiv \phi^{(s}_i(y,
\bar t_1,\bar t_2,...),\phi^{(s}_i\equiv \phi^{(s}_i(x,t_1,t_2,...) $
on evolution times parameters and space coordinates of the problem $x,y$. 

As a result we obtain the integrable hierarchies of evolution
type equations (each system of which determines by different
choice of $d_1,d_2$ under the fixed $m_1,m_2$) all invariant
with respect to UToda $(m_1,m_2)$ substitution.

In one-dimensional case $(\frac{\partial}{\partial y}=\frac{\partial}
{\partial y})$ this construction is equivalent to multi time formalism
with corresponding technique of Hamiltonians flows \cite{4}

Now we want to enumerate the problems which may be resolved in
context of the results of the present paper.

There is no doubts in possibility of the direct (literally)
generalization of this construction on supersymmetrical case.
As a result it will be possible to obtain unknown up to now integrable
hierarchies in $(2|2)$ superspace.

Excellent interesting is the problem of generalization on the
quantum domain. Heisenberg operators of usual two-dimentional
Toda chain (under canonical rules of quantization) \cite{12} may be
expressed as the matrix elements of single quantum group element
$k=m_-^{-1}m_+$, where elements of quantum groups $m_{\pm}$ are the
solution of the following system of equations
\begin{equation}
(m_+)_y=m_+(\sum_{s=1}^r X^+_s \exp (k\bar \phi(y))_s,\quad
(m_-)_x=m_-(\sum_{s=1}^r X^-_s \exp (k\phi(x))_s \label{03}
\end{equation}
$X^{\pm}_s$ now are the generators of the simple roots of quantum algebra,
$\phi(x)+\bar \phi(y)$ is the quantum solution of two dimensional
Laplace equation, $k$-Cartan matrix of semisimle algebra.

Comparison  (\ref{01}) with  (\ref{03}) shows us that the quantum
version of UToda chains is not a fantastic suggestion but the
problem of the nearest future. 

In the present paper we have considered only one example of the
system invariant with respect to UToda(2,2) substitution. We 
hope and shure that solution of symmetry equation in the case of
UToda substitutions may be solved by the similar methods as it was done
in \cite{3} in the case of the usual Toda chain. But now we are
not ready to solve this problem.

And the last comment. Reader can marked the deep disconnection
between the simplicity and pure algebraic nature of foundation of
construction (only single group element $k$ and equations (\ref{01}),
(\ref{02}) its define) from one hand and numerous nontrivial recurrent
relations between the matrix elements of the different fundamental
represention which it is necessary to prove by independent consideration
(under approach of the present paper) from the other side. It is
possible to hope that the last recurrent relations are the
direct corollary of (\ref{01}), (\ref{02}) but how to extract from
them this information is unknown to the author and may be this
is the most interesting problem for the future investigation and
the most important output of the present paper.

\section{Appendix}

Let us consider the function
$$
R^i=\langle i\parallel  K  X^+_{i-1} X^+_i \parallel i\rangle 
[ i-1 ]+\langle i-1\parallel  K X^+_iX^+_
{i-1} \parallel i-1\rangle [ i ]-
$$
$$
\langle i\parallel  K  X^+_i \parallel i\rangle \langle i-1\parallel  K X^+_
{i-1} \parallel i-1\rangle
$$
where $K$ is arbitrary group element of $SL(n,R)$ group.

From its definition and properties of the minimal state vector $\parallel i
\rangle$ (\ref{HV}) it follows that function $R$ is annihilated by
generators of all right positive and left negative simple roots.
This means that $R$ by itself is some linear combinations of the
matrixes elements taken between the minimal state vectors.
Calculations of left (right) Cartan elements shows that they
take the definite values on $R$ 
$$
h^s_r R^i=-(\delta_{s,i+1}+\delta_{s,i-2}) R^i,\quad h^s_l R^i=-(\delta_{s,i}+
\delta_{s,i-1}) R^i 
$$
From the last equalities it follows that from right and left $R^i$ 
belongs to different irreducible representation. This is
impossible and so $R^i=0$.

Now we enumerate the simplest state vectors of $i$ fundamental representation.
They are different by the number of generators of the simple
roots applied to the the minimal state vector.
Zero order $ \parallel i\rangle$ and first order $ X^+_i \parallel i\rangle$
states have the dimension one. Second order is two dimensional
$X^+_{i+1} X^+_i \parallel i\rangle,X^+_{i-1}X^+_i \parallel i\rangle$. 
There are three state vectors of the third order $X^+_{i+2}X^+_{i+1} X^+_i 
\parallel i\rangle,X^+_{i+1} X^+_{i-1} X^+_i \parallel i\rangle,X^+_{i-2}
X^+_{i-1} X^+_i \parallel i\rangle$ and fife ones of the
fourth order $X^+_{i+3}X^+_{i+2}X^+_{i+1} X^+_i \parallel i\rangle,
X^+_{i+2}X^+_{i-1}X^+_{i+1} X^+_i \parallel i\rangle,
X^+_i X^+_{i+1} X^+_{i-1} X^+_i \parallel i\rangle, 
X^+_{i-2}X^+_{i-1}X^+_{i+1}X^+_i \parallel i\rangle,
X^+_{i-3}X^+_{i-2}X^+_{i-1} X^+_i \parallel i\rangle$
All other possibilities give the state vectors with zero norm.

\end{document}